# Does cosmological vacuum energy density have an eletric reason ?

Claus W. Turtur
Fachhochschule Braunschweig- Wolfenbüttel
Salzdahlumer Straße 46 / 48
GERMANY - 38302 Wolfenbüttel
E-mail: c-w.turtur@fh-wolfenbuettel.de

**Abstract:**

Rather uncomplicated calculations by hand display a surprising connection between the energy density of the vacuum and the diameter of the universe. Among other things, the result explains the observation of the accelerated expansion of the universe.

**Structure of the Article:**

1. Explanation of the task
2. Calculation: The energy of the fields
3. Results: Energy density and diameter of the universe

## 1. Explanation of the Task

From cosmological investigations (see for instance [TEG 02]), we know that the universe consists of
- about 5 % normal matter, visible to mankind
- about 30 % cold dark matter (particles we can not see nowadays)
- about 65 % dark energy, also called vacuum energy

Obviously the major part of the energy in the universe is due to the vacuum. The visible consequence of this energy is its gravitation, following from the mass- energy- equivalence (E=mc²). Within the theory of relativity this gravitational effect is expressed by the cosmological constant $\Lambda$ (see for instance [GOE 96]). Measured proof of this gravitation is the acceleration of the expansion of the universe ([TON 03], [RIE 98]). Although gravitation should normally retard the expansion, measurement observes the opposite behaviour. A possible explanation could be given in this article here.

The nature of the vacuum energy is still under discussion:

- How did it come into the vacuum ?
- How is its consistency ?
- How can the vacuum store such amount of energy ?

It is likely, to assume an electric reason for this energy, this means an electric field. Anyhow we know, that also the 2.7 K cosmic background radiation is of electromagnetic nature. A background of an electric field would implicate, that the electric charge which causes this field originates from the big bang and since this time it is flowing into the space.

Considering the age of the universe of
$$T_0 \approx (16 \pm 4) \cdot 10^9 \, years = (5.05 \pm 1.26) \cdot 10^{17} \, \sec.$$ (for instance [PER 98], [RIE 98])
the electric charge today fills up sphere with a maximal diameter of
$$r_0 \approx c \cdot T_0 = (1.5 \pm 0.4) \cdot 10^{26} \, m$$ (with c = speed of light).
In this sense $r_0$ can be interpreted as diameter of the universe, which we regard as limited in contrary to the unlimited extension of the abstract mathematical space $\mathbb{R}^3$.

Remark: For the age and diameter of the universe, different values are known. Dynamic expansion of the universe leads to values different from the data achieved by radiological determination of the age. The above given value is estimated rather cautious, so that almost all of the values in literature are well within this interval.

The crucial question is now:
We regard a vacuum sphere, fulfilled with electrical charge, and we compare two forces. One force stems from the electric repulsion of the charge itself. The other force originates from the ponderable mass which is a result of the energy within the electric field. What we have to find is the diameter of the sphere in a way that both forces compensate each other exactly. This means, we are searching the diameter of a electrically charged vacuum-sphere, so that gravitational contraction and electrical expansion compensate each other.

The answer is:
In the following calculation we will see that the diameter of this sphere is just $r_0$, the diameter of our universe.

## 2. Calculation: The Energy of the Fields

This part of the article is addressed to the calculation of the absolute values of the energy, stored within the electric field and within the gravitational field of a sphere with the radius $r_0$ ([JAC 81]). We do not put any value of $r_0$ into the calculation, but we will receive a value for $r_0$ later in chapter 3.

Let us start with the definition of the symbols:
referring to the location in space:                                                                                        units:

$\rho_{elek}$ = local energy-density of the electrical field $\vec{E}$ $\rightarrow$ $\rho_{elek} = \frac{\varepsilon_o}{2} \cdot |\vec{E}|^2$   and   $[\rho_{elek}] = \frac{J}{m^3}$

$\rho_{grav}$ = local energy-density of the gravitational field $\vec{G}$ $\rightarrow$ $\rho_{grav} = \frac{1}{8\pi\gamma} \cdot |\vec{G}|^2$   and   $[\rho_{grav}] = \frac{J}{m^3}$ with $\varepsilon_o = 8.859 \cdot 10^{-12} \frac{C^2}{J \cdot m}$   and   $\gamma = 6.67 \cdot 10^{-11} \frac{N \cdot m^2}{kg^2}$         (equations 1a and 1b)

for the whole universe:

$\rho_M$ = mass-density of the vacuum (=universe) → units: $[\rho_M] = \frac{kg}{m^3}$ (numerical values: later)

$\rho_L$ = charge-density of the vacuum (=universe) → units: $[\rho_L] = \frac{C}{m^3}$

$\rho_U$ = energy-density of the vacuum (=universe) → $\rho_U = \rho_M \cdot c^2$, units: $[\rho_U] = \frac{J}{m^3}$

$W_{elek}$ = total energy of the electrical field of the sphere, units: $[W_{elek}] = J$

$W_{grav}$ = total energy of the gravitational field of the sphere, units: $[W_{grav}] = J$

Let us now find a mathematical expression for the gravitational field and let us calculate the energy-density resp. the energy of a sphere with the radius $r_0$.
This is not complicated, since we know from Isaac Newton the gravitational field of a sphere with homogeneously distributed mass (see figure 1 and equation 2).

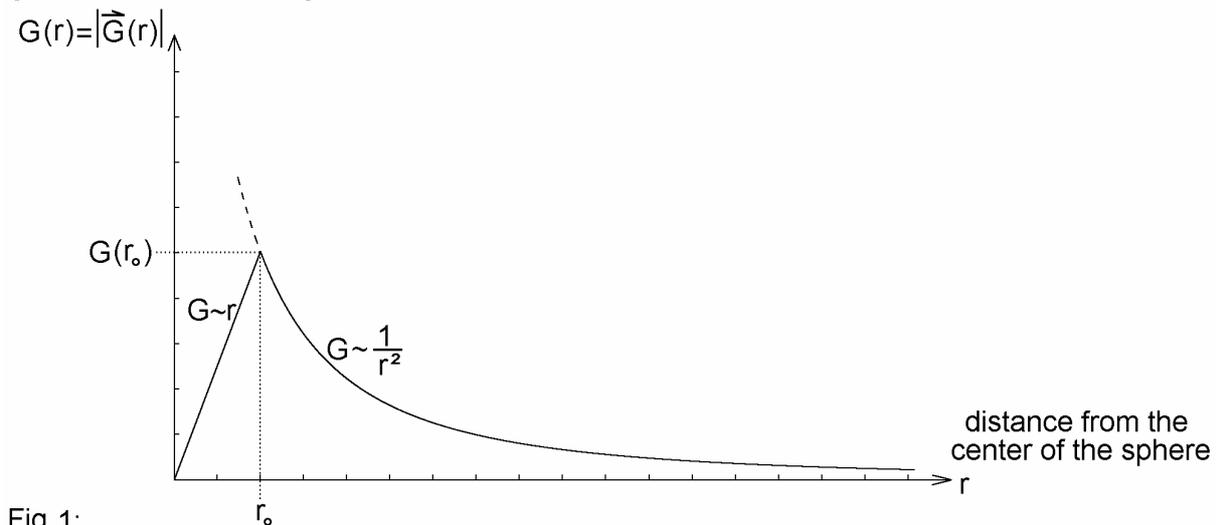

Fig.1:
Absolute value of the gravitational field strength of a homogeneous sphere with diameter $r_0$.

The mass within a spherical volume is: $m = \rho_M \cdot V = \rho_M \cdot \frac{4\pi}{3} r_0^3$

From this we derive the strength of the gravitational field at the distance from the center of $r_0$: $G(r_0) = \gamma \frac{\rho_M \cdot V}{r_0^2} = \frac{4\pi}{3} \gamma \rho_M r_0$

For the complete field strength as a function of r (=arbitrary distance from the center) we get:

$$G(r) = \begin{cases} \frac{4\pi}{3} \gamma \rho_M r & \text{for } 0 \leq r \leq r_0 \\ \frac{4\pi}{3} \gamma \rho_M \cdot \frac{r_0^3}{r^2} & \text{for } r \geq r_0 \end{cases}$$

(equation 2)

Because of the equivalence of all directions in the universe, we use spherical coordinates $r, \vartheta, \varphi$, hence the absolute value of the gravitational field strength does not depend on $\vartheta$ and $\varphi$.

From equations 1b and 2 the local value of this gravitational field strength is calculated as a function of the distance r from the center of the sphere:

$$\rho_{grav} = \frac{1}{8\pi\gamma} \cdot |\vec{G}|^2 = \frac{1}{8\pi\gamma} \cdot \begin{cases} \frac{16\pi^2}{9}\gamma^2 \rho_M^2 \, r^2 & \text{für } 0 \leq r \leq r_o \\ \frac{16\pi^2}{9}\gamma^2 \rho_M^2 \frac{r_o^6}{r^4} & \text{für } r \geq r_o \end{cases} = \frac{2\pi}{9}\gamma \rho_M^2 \cdot \begin{cases} r^2 & \text{für } 0 \leq r \leq r_o \\ \frac{r_o^6}{r^4} & \text{für } r \geq r_o \end{cases}$$

(equation 3)

The total energy of the field is calculated as usually by integration over the whole $\mathbb{R}^3$:

$$W_{grav} = \int_{\mathbb{R}^3} \rho_{grav} \, dV = \frac{2\pi}{9}\gamma\rho_M^2 \cdot \int_0^{2\pi}\int_0^\pi\int_0^{r_o} r^2 \cdot r^2 \sin\vartheta \, dr \, d\vartheta \, d\varphi + \frac{2\pi}{9}\gamma\rho_M^2 \cdot \int_0^{2\pi}\int_0^\pi\int_{r_o}^\infty \frac{r_o^6}{r^4} \cdot r^2 \sin\vartheta \, dr \, d\vartheta \, d\varphi$$

integration over $\vartheta$ and $\varphi$ only leads to a constant factor of $4\pi$.

$$= \frac{2\pi}{9}\gamma\rho_M^2 \cdot 4\pi \int_0^{r_o} r^4 \, dr + \frac{2\pi}{9}\gamma\rho_M^2 \cdot 4\pi \cdot \int_{r_o}^\infty \frac{r_o^6}{r^4} \, dr = \frac{16}{15}\pi^2 \gamma \rho_M^2 \cdot r_o^5$$

(equation 4)

This is the gravitational energy of the sphere.

The now following calculation of the electrical energy of the sphere is done analogously, because the charge distribution is the located analogous to the mass distribution:

The total electrical charge within the sphere is: $\quad Q = \int_{\mathbb{R}^3} \rho_L \, dV \quad$ (equation 5)

If $\vec{x}'$ is the location of the charge,
the electrical field strength at the location $\vec{x}$ will be: $\vec{E}(\vec{x}) = \frac{1}{4\pi\varepsilon_o} \cdot \int_{sphere} \rho_L \frac{\vec{x} - \vec{x}'}{|\vec{x} - \vec{x}'|^3} \, d^3x'$

Because of the homogeneity of the charge-distribution within the sphere, we derive:

$$\vec{E} = \begin{cases} \frac{Q}{4\pi\varepsilon_o} \cdot \frac{r}{r_o^3} & \text{für } 0 \leq r \leq r_o \\ \frac{Q}{4\pi\varepsilon_o} \cdot \frac{1}{r^2} & \text{für } r \geq r_o \end{cases} \implies \vec{E} = \frac{\rho_L}{3\varepsilon_o} \cdot \begin{cases} r & \text{für } 0 \leq r \leq r_o \\ \frac{r_o^3}{r^2} & \text{für } r \geq r_o \end{cases}$$

(equation 6)

Inserting $\rho_{elek} = \frac{\varepsilon_o}{2} \cdot |\vec{E}|^2$ in the case of the electrical field we get:

$$W_{elek} = \int_{\mathbb{R}^3} \rho_{elek} \, dV = \frac{\varepsilon_o}{2} \cdot \frac{\rho_L^2}{9\varepsilon_o^2} \cdot \int_0^{2\pi}\int_0^\pi\int_0^{r_o} r^2 \cdot r^2 \sin\vartheta \, dr \, d\vartheta \, d\varphi + \frac{\varepsilon_o}{2} \cdot \frac{\rho_L^2}{9\varepsilon_o^2} \cdot \int_0^{2\pi}\int_0^\pi\int_{r_o}^\infty \frac{r_o^6}{r^4} \cdot r^2 \sin\vartheta \, dr \, d\vartheta \, d\varphi$$

integration is performed in the same way as in equation 4

$$= \frac{\rho_L^2}{18\varepsilon_o} \cdot 4\pi \int_0^{r_o} r^4 \, dr + \frac{\rho_L^2}{18\varepsilon_o} \cdot 4\pi \cdot \int_{r_o}^\infty \frac{r_o^6}{r^4} \, dr = \frac{4\pi}{15\varepsilon_o} \cdot \rho_L^2 \cdot r_o^5$$

(equation 7)

The calculation of the energy within the fields is now done. These energies are known under the name self-energy. We should keep in mind, that we only calculated the absolute values of these energies, not taking the algebraic signs into account. In fact gravitation is an attractive force, but Coulomb-force is repulsive. If gravitation dominates, the sphere will contract; if Coulomb-force dominates, the sphere will expand.

If we want to find the equilibrium of the forces, we have to adapt equation 4 und equation 7 to each other, because the charge density $\rho_L$ and the mass density $\rho_M$ are different physical expressions. Fortunately these both densities are connected to each other in a rather simple way, because the ponderable mass is a result of the energy within the Coulomb- field according to Einstein's energy- mass- equivalence (W=mc²). On this background we can do the conversion as following:

The mass is: $M = W_{elek} \cdot c^2 \implies$ The density of the mass is $\rho_M = \dfrac{M}{V} = \dfrac{W_{elek}}{V \cdot c^2}$

with equation 7 we derive $\rho_M = \dfrac{\frac{4\pi}{15\varepsilon_o} \cdot \rho_L^2 \cdot r_o^5}{\frac{4\pi}{3} r_o^3 \cdot c^2} = \dfrac{\rho_L^2 \cdot r_o^5}{5\varepsilon_o \cdot r_o^3 \cdot c^2} \implies \rho_L^2 \cdot r_o^5 = \rho_M \cdot 5\varepsilon_o \cdot r_o^3 \cdot c^2$  (equation 8a) unsing again equation 7 we come to the required conversion $\implies W_{elek} = \dfrac{4\pi}{15\varepsilon_o} \cdot \rho_L^2 \cdot r_o^5 = \dfrac{4\pi}{15\varepsilon_o} \cdot \rho_M \cdot 5\varepsilon_o \cdot r_o^3 \cdot c^2 = \dfrac{4\pi}{3} \cdot \rho_M \cdot c^2 \cdot r_o^3$  (equation 8b)

With the equations 4 and 8b we reached our aim to find expressions for the electrical and the gravitational energy of the sphere, that can be compared mathematically. This comparison will be the subject of chapter 3 in this article.

## 3. Results: Energy density and diameter of the universe

Summation of the electric and the gravitational energy (according to the equations 4 and 8b) allows the determination of the total energy of the sphere, we just have to use to correct algebraic sign to each expression – see equation 9. Gravitation is attractive (negative sign), but Coulomb- forces are repulsive (positive sign).

$$W_{tot} = W_{elek} - W_{grav} = \dfrac{4\pi}{3} \cdot \rho_M \cdot c^2 \cdot r_o^3 - \dfrac{16}{15}\pi^2 \gamma \rho_M^2 \cdot r_o^5 \qquad \text{(equation 9)}$$

The location of the equilibrium is to be found at minimum of the total energy:

$$\dfrac{d}{dr} W_{tot} = 4\pi \cdot \rho_M \cdot c^2 \cdot r_o^2 - \dfrac{16}{3}\pi^2 \gamma \rho_M^2 \cdot r_o^4 = 0$$

(only the positive square-root is senseful)

solving to $r_o \implies r_o = \pm \sqrt{\dfrac{3c^2}{4\pi \gamma \rho_M}} = (1.8 \pm 0.3) \cdot 10^{26}\, m = (19 \pm 3) \cdot 10^9$ light- years  (equation 10) with a vacuum- energy- density of $\rho_M = (1.0 \pm 0.3) \cdot 10^{-26}\, \dfrac{kg}{m^3}$ (reason is following soon)

The reason fort he value of the vacuum- energy- density is:
[GIU 00] points out a value of about $\rho_M \approx 10^{26}\, \frac{kg}{m^3}$.
In [TEG 02] (his page no.3) we find a "Vacuum density constant" with theoretical background of $\rho_M \approx 1.15 \cdot 10^{26}\, \frac{kg}{m^3}$ and a "Matter density" coming from measurement of $\rho_M \approx 0.75 \cdot 10^{26}\, \frac{kg}{m^3}$. Other different sources give different values for $\rho_M$ which contain fluctuations to somehow larger values. From this point of view, let us decide to use a value of $\rho_M \approx (1.0 \pm 0.3) \cdot 10^{26} \frac{kg}{m^3}$ (Gleichung 11)

But one position after decimal point can be regarded to be enough. Probably the given uncertainty interval is a bit optimistic.

Calculus of uncertainty for $r_0$ in equation 10 in principle is performed as Gaussian error propagation but with only one erroneous variable $\rho_M$.

The result of equation 10 is consistent with the age of the universe within the interval of uncertainty. From this we conclude:

**The article presents a rather simple model for a theoretical derivation of the age and the diameter of the universe. The only necessary assumption is the energy density of the vacuum. The results of the model suit surprisingly good with the well-known values in literature.**

If we apply equation 10 the opposite way around, we can put $r_0 \approx (1.5 \pm 0.4) \cdot 10^{26} m$ into the calculation and derive the mass- density of the universe. We find a value of:

$$\rho_M = \frac{3c^2}{4\pi \gamma r_0^2} = (1.43 \pm 0.76) \cdot 10^{-26} \frac{kg}{m^3} \quad \text{(equation 12)}$$

Calculus of uncertainty is again performed as Gaussian error propagation with only one erroneous variable (now $r_0$). The interval of uncertainty is less optimistic than in equation 11, but it could be realistic in a way.

**So the given value for the energy- density of the universe should certainly be regarded as reasonable. The only assumption for the calculation is the age of the universe.**

For the benefit of interest, the density of electrical charge within the vacuum (of the universe) can be calculated as following (values for $\rho_M$ and $r_0$ are taken from equation 10):

Frorm equation 8a we derive: $\rho_L = \sqrt{\rho_M \cdot 5\varepsilon_0 \cdot c^2} \cdot \frac{1}{r_0} = (1.11 \pm 0.38) \cdot 10^{-36} \frac{C}{m^3}$ (equation 13)

Concerning technical utilization, this value is not very great – but may one day the electric charge of the vacuum can be verified and technically used to extract energy from the vacuum ?

Here are some final remarks, which might be additionally relevant:

(a.) The calculation assumes the homogeneity of the universe, this means, that the vacuum has the same density everywhere in the universe. The assumption can be under doubt (see for instance [GIU 00]). A reasons could be, that different parts of the universe escape with different velocity from the point of the big bang. The inhomogeneity is not a problem for our model, but it would only alter the integration of the equations 4 and 7.

(b.) Up to now our model includes only gravitation and Coulomb- forces. In principle it would be possible to expand equation 9 to further forces, for instance to all fundamental interactions, such as strong and weak interaction. Nevertheless, strong and weak interaction should play only a secondary role, because they are of rather short range and can only affect the space very close to particles with rest mass. Also from the energy- percentage quoted in chapter 1 we know, that they are not dominant for the average over the whole universe. But perhaps from their small contribution to the energy of the whole universe, the small differences in the vacuum- mass- density between equation 11 and equation 12 can be explained, as well as the small differences for the size and the age of the universe between equation 10 and the value in chapter 1.

(c.) So what – does our model explain an acceleration of the expansion of the universe or a deceleration ?   BOTH !
The point is: Up to here, we performed a calculation of the forces' equilibrium between gravitation and Coulomb- force. But we did not take inertia of masses into account.
If we assume, that in the moment of the big bang, electric repulsion did dominate extremely strong, the universe moved towards the equilibrium position in the very beginning. Because of masses' inertia the movement of the universe came to an oscillation around the equilibrium position. If the oscillation would not be attenuated, the universe would come back to the point of the big bang one day. But on the other hand, if we have attenuation within the universe, the amplitude would decrease and the oscillation might happen nearby the equilibrium position today. It could be rather likely, we reached this possible state nowadays – many billions of years after the big bange.
In this case, if the universe oscillates around the equilibrium position with not too large amplitudes, it will alternately expand and contract itself within billions of years. This would be a reasonable explanation for those cosmological measurements we observe today.
Besides these possibilities, we could have the option in mind, that some exothermal reactions take place in the universe which meanwhile begin to keep the amplitude constant in time, or even to enlarge it. This would not alter the equilibrium position, but it could bring back the chance for a collapse one day.